\documentclass[%
 aip,
 amsmath,amssymb,
 reprint,%
]{revtex4-1}

\usepackage{graphicx}
\usepackage{dcolumn}
\usepackage{bm}

\usepackage[utf8]{inputenc}
\usepackage[T1]{fontenc}
\usepackage{mathptmx}

\begin{document}


\title[]{2D-subband spectra variations under persistent tunnelling
photoconductivity condition in tunnel delta-GaAs/Al structures}

\author{S.~E.~Dizhur}
\email{sdizhur@cplire.ru}
\affiliation{Institute of Radioengineering and Electronics RAS, 101999 Moscow, Russia}

\author{I.~N.~Kotel'nikov}
\affiliation{Institute of Radioengineering and Electronics RAS, 101999 Moscow, Russia}

\author{V.~A.~Kokin}
\affiliation{Institute of Radioengineering and Electronics RAS, 101999 Moscow, Russia}

\author{F.~V.~Shtrom}
\affiliation{Institute of Radioengineering and Electronics RAS, 101999 Moscow, Russia}


\begin{abstract}
2DEG states of Al/$\delta$-GaAs structures were investigated in the persistent tunnelling photoconductivity (PTPC) regime at low temperatures. 
''Thickening'' of the unoccupied subbands to the ground state of 2DEG was observed at $T=4.2$~$K$.
It was found that there is a uniform shift of the subband levels at $T=77$~$K$.
The behaviour of the persistent 2D states in tunnelling spectra after various illumination and at different temperatures allows us to assume that there are of two PTPC mechanisms, namely, the broadening of the
potential well profile of the $\delta$-layer and photoionization of the DX-centers.
\end{abstract}

\maketitle

\section{\label{sec:Intro}Introduction}

We began to take  interest in persistent tunnelling photoconductivity effect during experiments concerning the intersubband resonant polaron in the Al/$\delta$-GaAs structures \cite{1}, in which the diamagnetic shift of the 2D subbands in the $\delta$-layer was used.
The positions of the 2D subbands can be changed by persistent photoconductivity (PPC) too.
This effect well known for the case of occupied subbands of the $\delta$-layers in the bulk of GaAs (see, for instance, \cite{2a,2}).
The concentration $n_{2D}$ of the two dimensional electron gas (2DEG) increases after the pulse illumination of such structures at liquid helium temperatures.
Occupations of the 2D subbands in the PPC regime are determined from magnetotransport measurements.
The origin of the increase of $n_{2D}$ of the $\delta$-doped structures in the PPC regime is not clear yet.
This effect is explained by the influence of the DX-centers, interface states, deep levels of the GaAs gap, etc \cite{3,4,4a,5}.
In recent years communications about new features of the PPC in $\delta$-GaAs at low-doping densities \cite{6a} and about the effects of charge correlations in the samples of $\delta$-GaAlAs \cite{6} appear.
As far as we know, the investigations of PPC in the near-surface $\delta$-layers ($\gtrsim 30$~$nm$ from
GaAs surface) were conducted by the capacitance-voltage profiles technique and by the deep level transient spectroscopy only (see, e.g. \cite{3}).
But such methods do not allow to analyze the energy spectrum of 2DEG in the $\delta$-layers.
Though the near-surface $\delta$-layer can be characterized by tunnelling spectroscopy \cite{1}, these experiments in PPC regime were not carried out.
The tunnelling spectroscopy, as distinct from the traditional magnetotransport measurements, allows to obtain spectra of both occupied and unoccupied 2D subbands directly.
Therefore the modification of the self-consistent potential profile near $\delta$-doped layer can be studied under the PPC conditions.
The first persistent tunnelling photoconductivity experiments in the Al/$\delta$-GaAs structures demonstrated that a significant variation of the 2D subband spectrum occurs in the unoccupied subbands \cite{7}.

\section{Samples and experiment}

\subsection{Samples}

\begin{figure}[ht]                                                                                   
\includegraphics[width=\columnwidth]{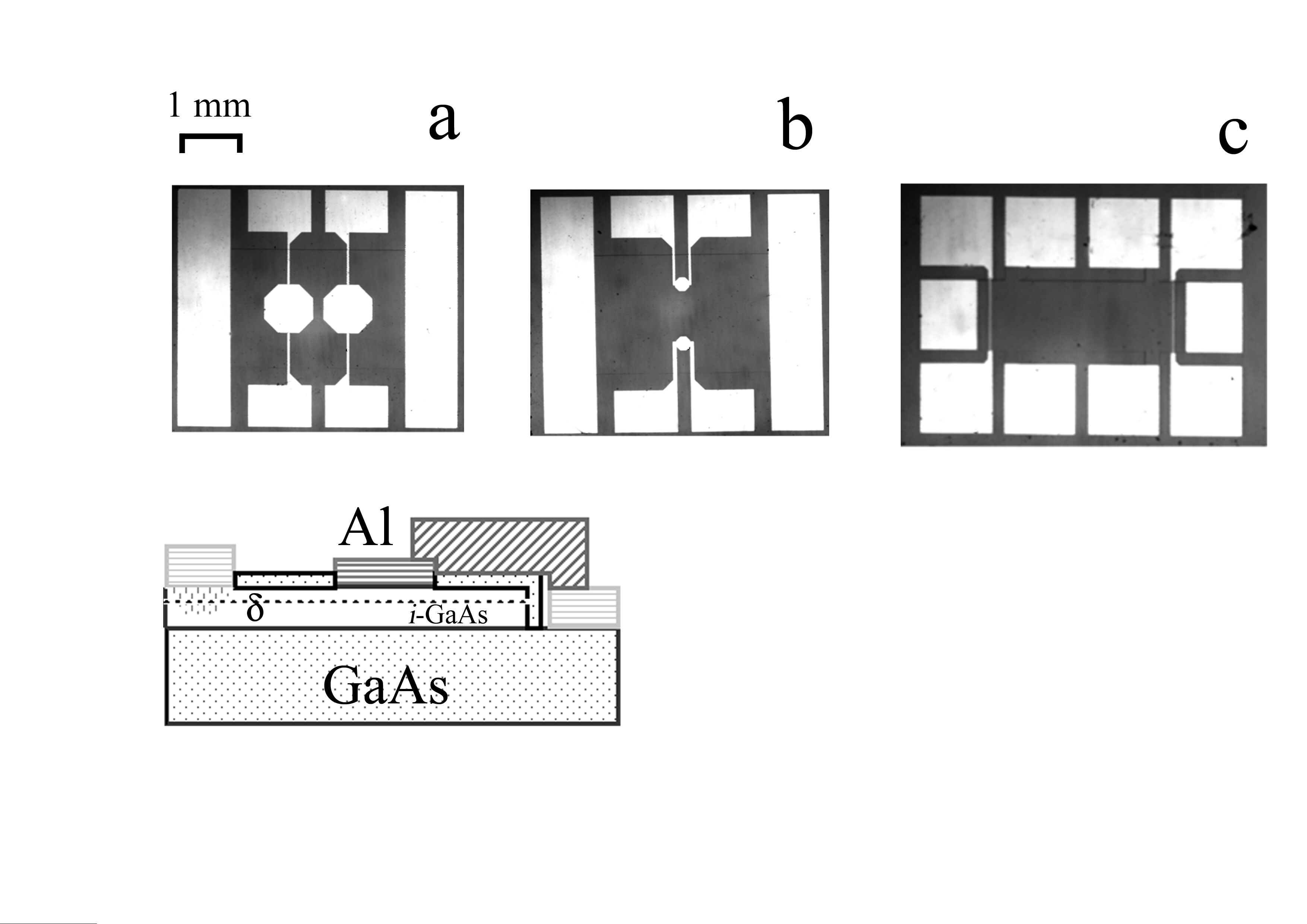}
\caption{
Microphotographs of the different samples 'a', 'b', 'c' and sketchy sectional view of Al/$\delta$-GaAs structure.
The dimensions of the octagonal Al electrodes of the samples 'a' and 'b' are 0.7 and 0.2~$mm$ respectively, the width of Al strips of the sample 'c' is 10~$\mu m$.}
\label{photo}
\end{figure}

\begin{figure}[ht]
\includegraphics[width=\columnwidth]{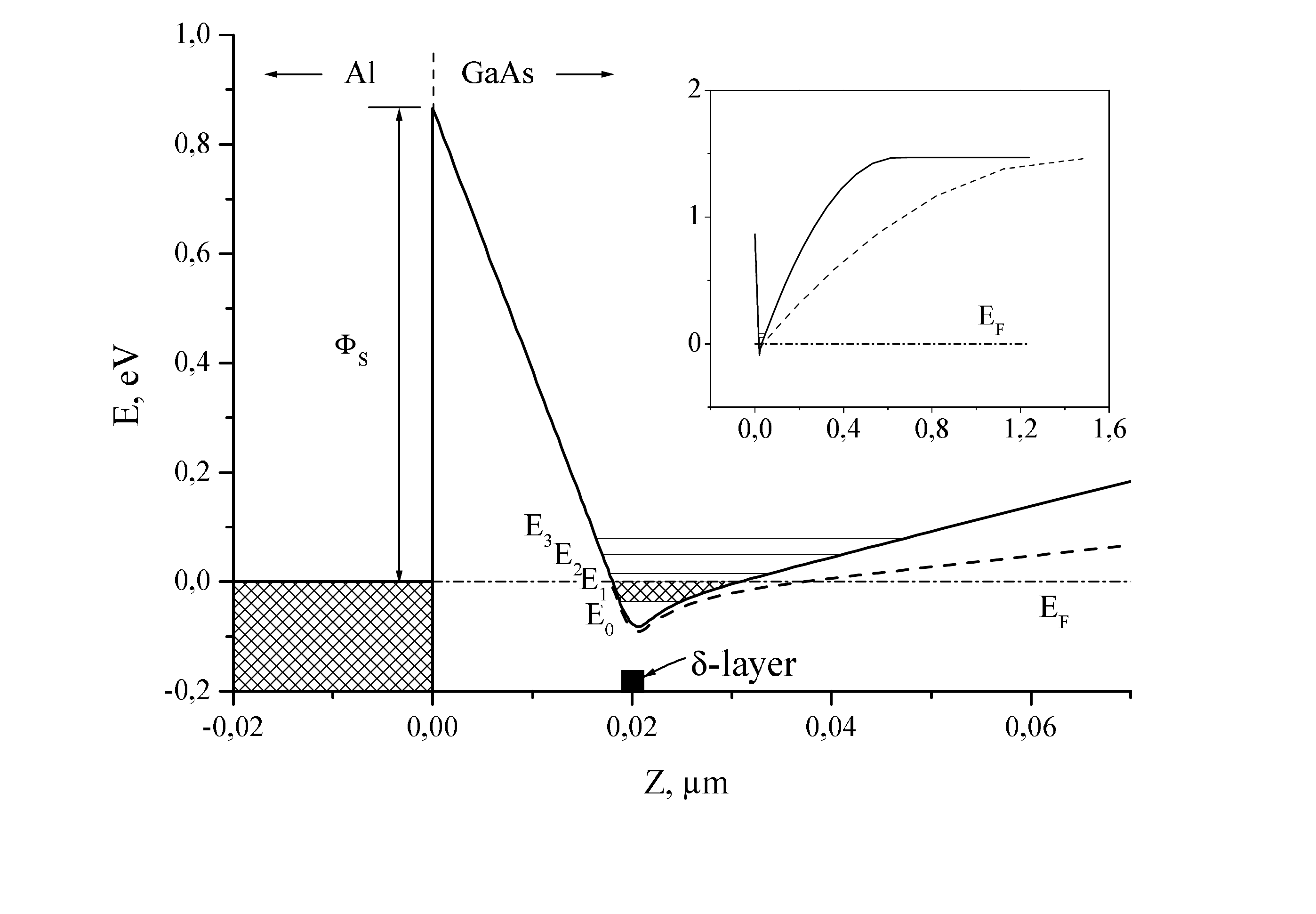}
\caption{
Band diagram of $\delta$-doped structures. 
Self-consistent calculation of the potential well profile was made with the following
parameters (best approximation for the samples d9): donor (Si) concentration in $\delta$-layer
$N^{\delta}_{Si}=4.7\times10^{12}$~$cm^{-2}$; distance between $\delta$-layer and Al/GaAs interface $z_{\delta}=20$~$nm$; $\delta$-layer width $\Delta z_{\delta}=2.5$~$nm$; $\Phi_{S}=0.87$~$eV$; bias voltage at tunnel junction $U=0$; acceptor concentrations in GaAs $N_{a}=5.5\times10^{15}$ and $1\times10^{15}$~$cm^{-3}$ (dashed line). 
The insert shows potential profile in the bulk of GaAs.}
\label{qw}
\end{figure}

The tunnel structures Al/$\delta$-GaAs were prepared in IRE RAS on semi-insulating (100) GaAs substrates by the molecular beam epitaxy (MBE) method.
The $\delta$-doped layer with the density of Si atoms $5.2\times 10^{12}$~$cm^{-2}$ was formed at
$570~{^{\circ}}C$ at the distance of $L\simeq 20$~$nm$ from Al/GaAs interface.
Deposition of Al from the Knudsen cell took place directly in the MBE chamber \cite{1}.
The width of a semitransparent Al film was 80--100~$nm$.
Al/$\delta$-GaAs tunnel junctions were photolitographically defined and etched into samples.
The ohmic contacts to the $\delta$-layer were made by Au-Ge-Ni alloy.
The sketchy sectional view of this structure is shown at Figure~\ref{photo}.
Three different geometric of samples ('a', 'b', 'c') were used in our measurements  as shown in the same figure.
The charge of the impurity of the $\delta$-layer forms a strong asymmetric (due to its closeness to the interface) V-shaped potential well with 2DEG.
2D electron sheet concentration in our samples was in the range of $0.5\times10^{12}$ to $1.5\times10^{12}$~$cm^{-2}$, as it was found from tunnel spectra (TS);
2DEG mobility was equal approx to 1500~$cm^{2}/(Vs)$. 
The band diagram of such QW is shown on Figure~\ref{qw}, where the potential profile was calculated
self-consistently. Here $\Phi_{S}$ is a barrier height at Al/GaAs interface; $E_{F}$ -- Fermi energy in GaAs; $E_{0}, E_{1} \ldots$ -- subbands levels in $\delta$-layer QW.

\subsection{Experiment}

\begin{figure}[ht]
\includegraphics[width=\columnwidth]{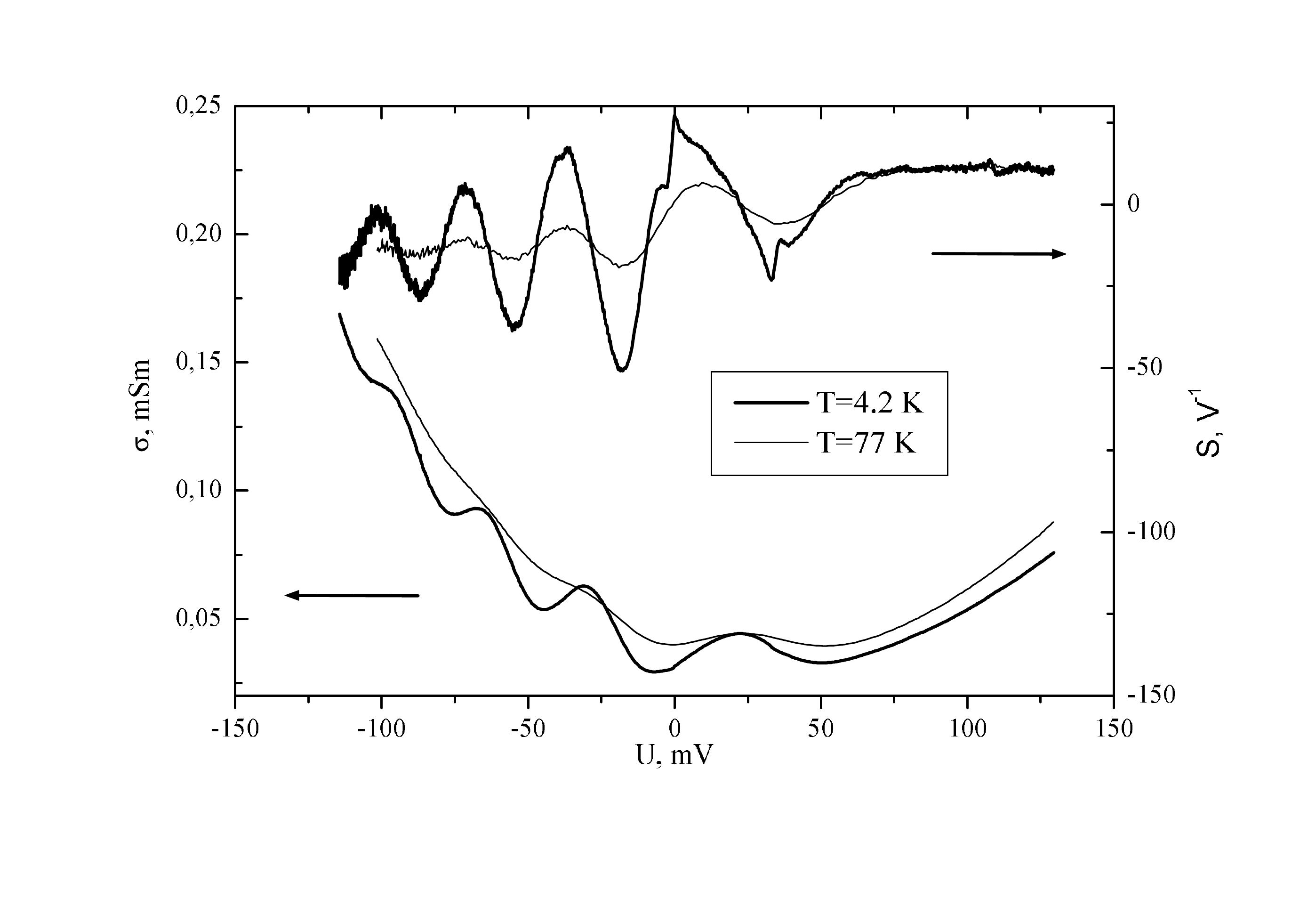}
\caption{
Tunnel conductance $\sigma$ (bottom curves and left axis) and its logarithmic derivative $S=d\ln\sigma/dU$ (top curves and right axis) at temperatures $T=4.2$ and 77~$K$ for sample d9a11 ($n_{2D}=1.1\times 10^{12}$~$cm^{-2}$).}
\label{ssnh}
\end{figure}

The experiments were made at temperatures from 4.2 to 300~$K$.
TS was obtained by standard lock-in technique using the 4-contact method of measurement.
Two lock-in amplifiers simultaneously detected first and second harmonics of the signal, proportional to
$dU/dI$  and $d^{2}U/dI^{2}$ of the U-I curve respectively. 
Tunnel spectrum $S(U)=d\ln\sigma/dU$, where $\sigma=dI/dU$, was calculated from experimental data.
Tunnel conductance $\sigma(U)$ and tunnel spectrum $S(U)$ at temperatures $T=4.2$ and 77~$K$ are shown in Figure~\ref{ssnh}.
Similar curves were obtained for different geometric of the samples (Figure~\ref{photo}) at the same $n_{2D}$.
We make use of tunnel spectra in $S(U)$ form, because of convenience from several points of view.
First, this value is independent of absolute magnitude of conductance, that allow us to compare TS of the different metal electrodes diameters.
Second, $S(U)$ is measured in universal units of $V^{-1}$.
When conductance is the exponential function of the bias voltage $U$ then $S(U)=const$.
For our junctions Al/$\delta$-GaAs potential barrier is extremely close to linear function of the coordinate (see Figure~\ref{qw}).
It is easy to show \cite{1} (for quasi-classical approximation and one occupied subband energy $E_{0}$) that at bias voltages $U \simeq +100$~$mV \ll (\Phi_{S}+E_{F})/e \simeq 1$~$V$ the tunnel conductance
$$
\sigma \sim D(E_{0}, eU) \sim \exp\left(\frac{eU-2(\Phi_{S}+E_{F})}{\mathcal{E}_{00}}\right),
$$ 
where 
$$\mathcal{E}_{00}=\frac{3}{2}\frac{\hbar}{\sqrt{2m}}\frac{\sqrt{\Phi_{S}+E_{F}}}{z_{\delta}}$$.

Figure~\ref{ssnh} shows that such a simplified model describes well TS behaviour at large positive $U$.
The valuation $e/\mathcal{E}_{00}\simeq 14$~$V^{-1}$ is easy determine from the shown spectra in the region $U\simeq 100$~$mV$, and gives quantity of the characteristic energy $\mathcal{E}_{00}\simeq 71$~$meV$.
The calculation by formula above gives good agreement with the experiment for parameters of junction $\Phi_{S}+E_{F} = 1.1$~$eV$ and $z_{\delta}=17$~$nm$.
Each strong dip in TS is related to the bottom $E_{i}$ of the two-dimensional subbands in the ${\delta}$-layer.
Tunnel conductance in 2DEG tunnel structure is combined of the conductance of 2D subbands:
$\sigma(U)=\sum_{i}\sigma_{i}\propto\sum_{i}\rho_{\|i}(E_{i}-eU)\times D(E_{i}, E_{F}-eU)$, where $i$ is the subband's number, $\rho_{\|i}=(m/\pi\hbar^{2})\times\Theta(E_{F}-E_{i}-eU)$ is the two-dimensional density of states (2DOS), $D$ is the barrier transmission, $E_{F}$ is Fermi energy and $U$ is bias voltage on
the tunnel junction; 
function $\Theta$, by definition, is $\Theta \equiv 1$ for $x \geq 0$ and zero otherwise.
The derivation of the stepped 2DOS gives the dips seen in TS.
Thus the positions $U_{i}$ of the dips in $d\ln\sigma/dU$ give the subband energies $E_{i}$ in 2DEG (with the accuracy of weak dependance $E_{i}(U)$ \cite{1}).
Figure~\ref{ssnh} shows that the amplitudes of the dips decrease at $T=77$~$K$. It is associated with degradation of the tunnel spectroscopy resolution with increasing temperature.
But the positions $U_{i}$ of the dips are still rather well measured at $T=77$~$K$.
Note that the dips at $U>0$ and $U<0$ correspond to occupied and unoccupied subbands, respectively.
The many-body features in TS at 4.2~$K$ (near $U=0$ and $U=\hbar \omega_{LO}/e=36.5$~$mV$) will not be discussed in this paper.

In PTPC experiments, the samples were cooled to the temperature 4.2~$K$ under dark conditions, then they were illuminated by the pulses of the red LED mounted in cryostat ($h\nu>E_{g}$, where $E_{g}$ is GaAs energy gap).

\section{Results and discussion}

\begin{figure}[ht]
\includegraphics[width=\columnwidth]{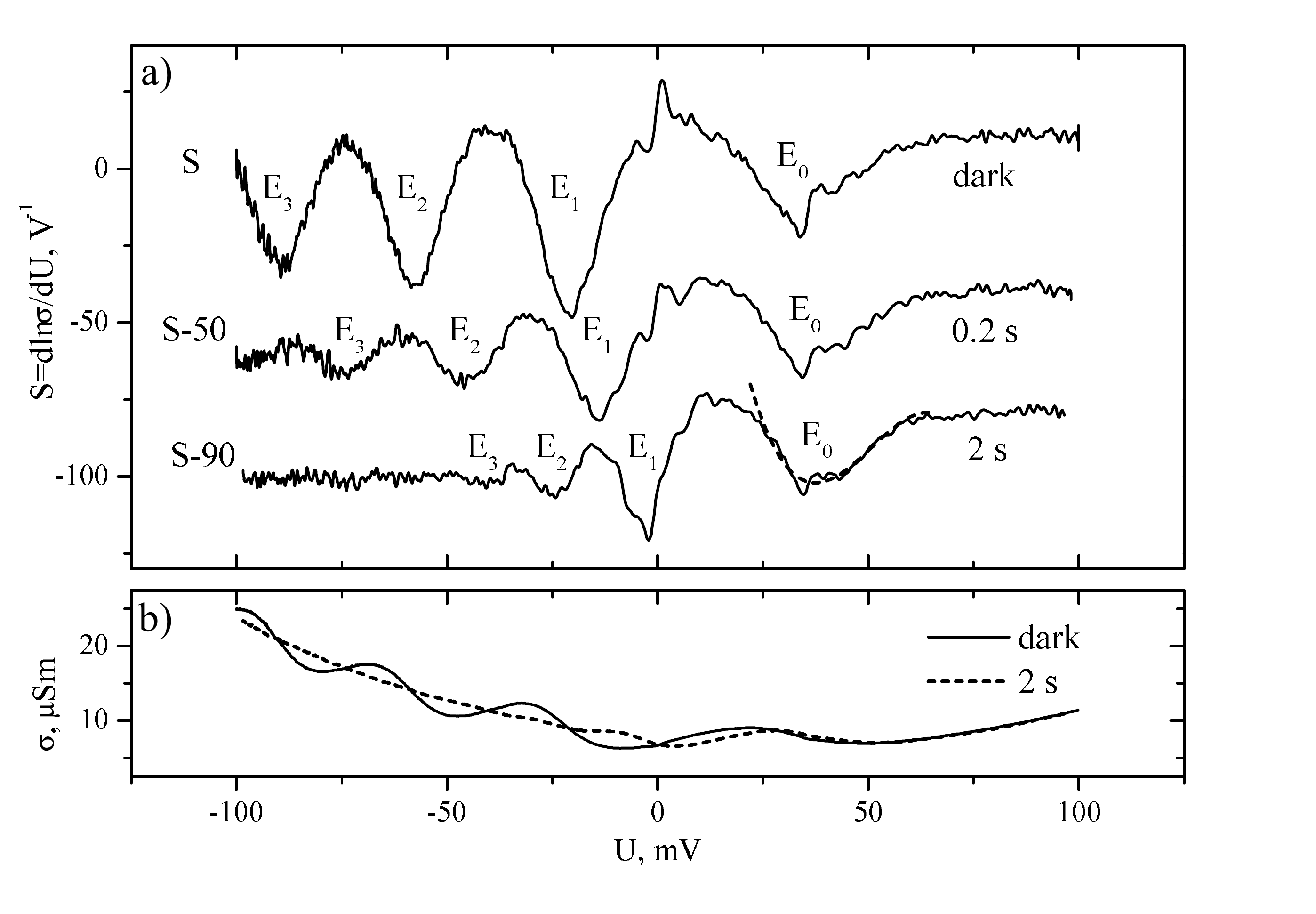}
\caption{
a) tunnel spectra at the dark (top curve) and different illuminations (bottom curves). Signs $E_{i}$ show subbands positions.
Spectra at illumination time $t=0.2$ and 2~$s$ are shifted from dark spectrum at 50 and 90~$V^{-1}$.
Dashed line on the bottom curve shows polynomial interpolation of one of the subbands.
b) tunnel conductance at the dark (solid line) and after illumination time $t=2$~$s$ (dash line).
Temperature is 4.2~$K$. Sample d9b91.}
\label{s_l}
\end{figure}

Figure~\ref{s_l} demonstrates TS of the junction Al/$\delta$-GaAs with concentration $n_{2D}=1.1\times 10^{12}$~$cm^{-2}$ at $T=4.2$~$K$ after cooling under dark conditions (upper curve) and different
illuminations (bottom curves).
According to Figure~\ref{s_l},  only one subband $E_{0}$ is occupied in this sample and the value of Fermi energy is $eU_{0}=-E_{0}\approx 40$~$meV$.
In all tested samples only $E_{0}$ subband was partially filled under dark conditions and the
position $U_{0}$ in TS at different 2D electron sheet concentrations was changed from 20 to 50~$mV$.
Illumination of the samples by LED at liquid helium temperatures in time $t$ led to variation of the tunnel conductance and of the dips positions in TS (bottom curves on Figure~\ref{s_l}) \cite{7}.
It is well seen from Figure~\ref{s_l} that the main consequence of illumination is the change of the 2D subband spectra, instead of the conductance change.
Observed PTPC and subband spectra variation depend upon ''dark'' concentration of 2D electrons and reaches saturation that corresponds to the illumination time.
The persistent state of the illuminated tunnel structure was not actually changed in several hours: for example, the tunnel conductance decreased for less than 0.1\% from the initial value in one
hour. 
But nevertheless, the dip positions were not changed within the accuracy of our measurements.

\begin{figure}[ht]
\includegraphics[width=\columnwidth]{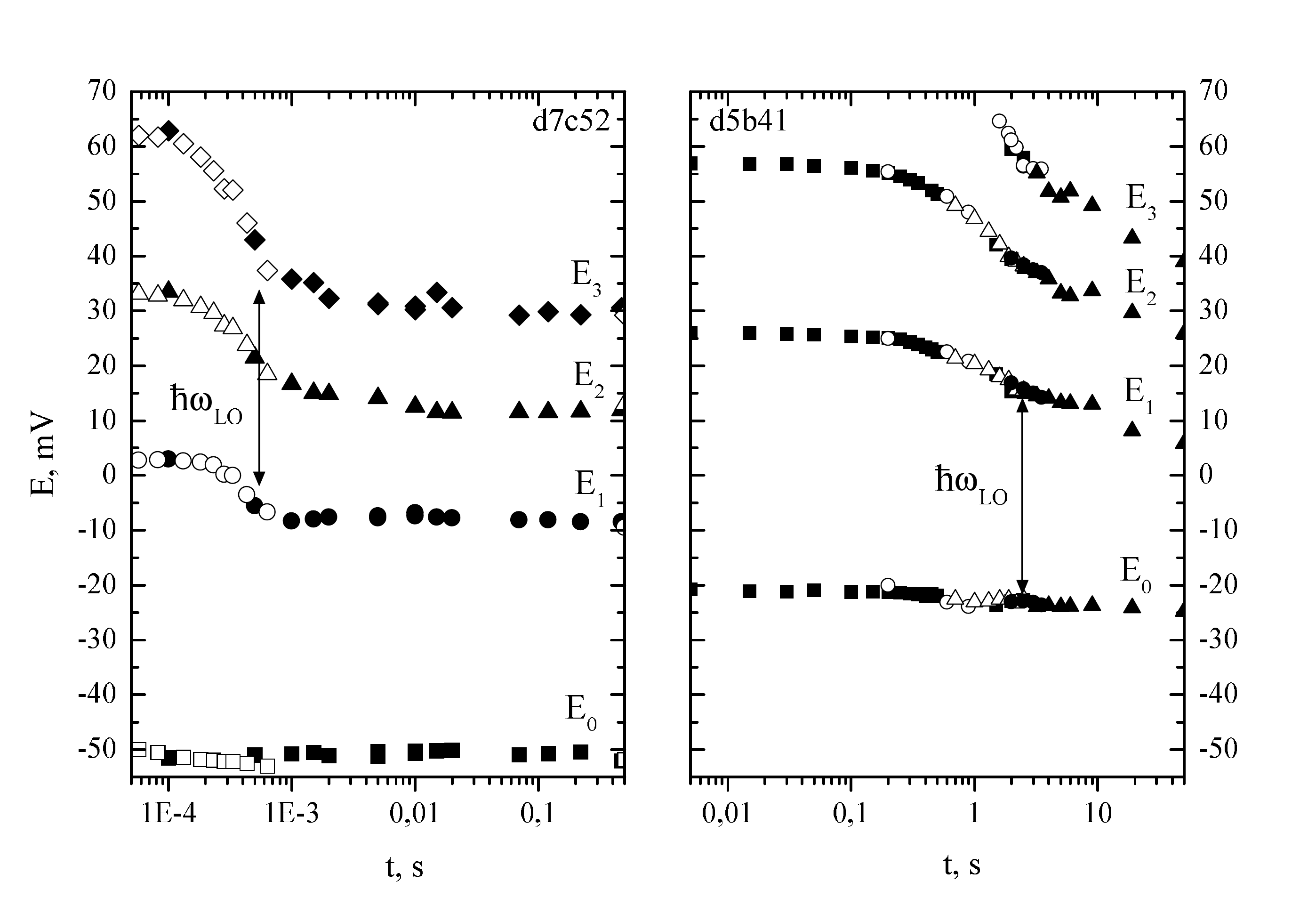}
\caption{
Subband energies $E_{i}$ of the samples d7c52 and d5b41 as functions of the illumination time $t$ at $T=4.2$~$K$.
The thicknesses of Al films of these junctions were 80 and 100~$nm$ respectively.
Different signs corresponds to different series of experiments. 
$\varepsilon_{LO}=\hbar\omega_{LO}$ is the LO-phonon energy.}
\label{e_l_2}
\end{figure}

Heating of the sample to room temperature with following cooling in the dark returned the structures parameters to the initial (''dark'') values.
The ground state position weakly depends on $t$ that corresponds to magnetotransport measurements for the occupied subbands of the ''bulk'' $\delta$-layers at PPC regime.
The difference $E_{0}(0)-E_{0}(t)$ at saturation corresponds to increase of 2D electron concentration at 10--15\%.
Figure~\ref{e_l_2} demonstrates subbands energies (measured from $E_{F}$) as functions of $t$, calculated from dip positions in TS.
For more precise determination of minima of the dips, the approximation with polynomial of power 3--5 was used (see,  for instance, the dashed curve for $E_{0}$ on Figure~\ref{s_l}).
It is seen that for empty subbands the shifts are 10--50~$meV$ and for filled level $E_{0}$ shift is 2--5~$meV$.
Note that at some $t$ on Figure~\ref{e_l_2} the resonant conditions for intersubband polaron interaction are realized too.
We will discuss theses effects in future papers.

\begin{figure}[ht]
\includegraphics[width=\columnwidth]{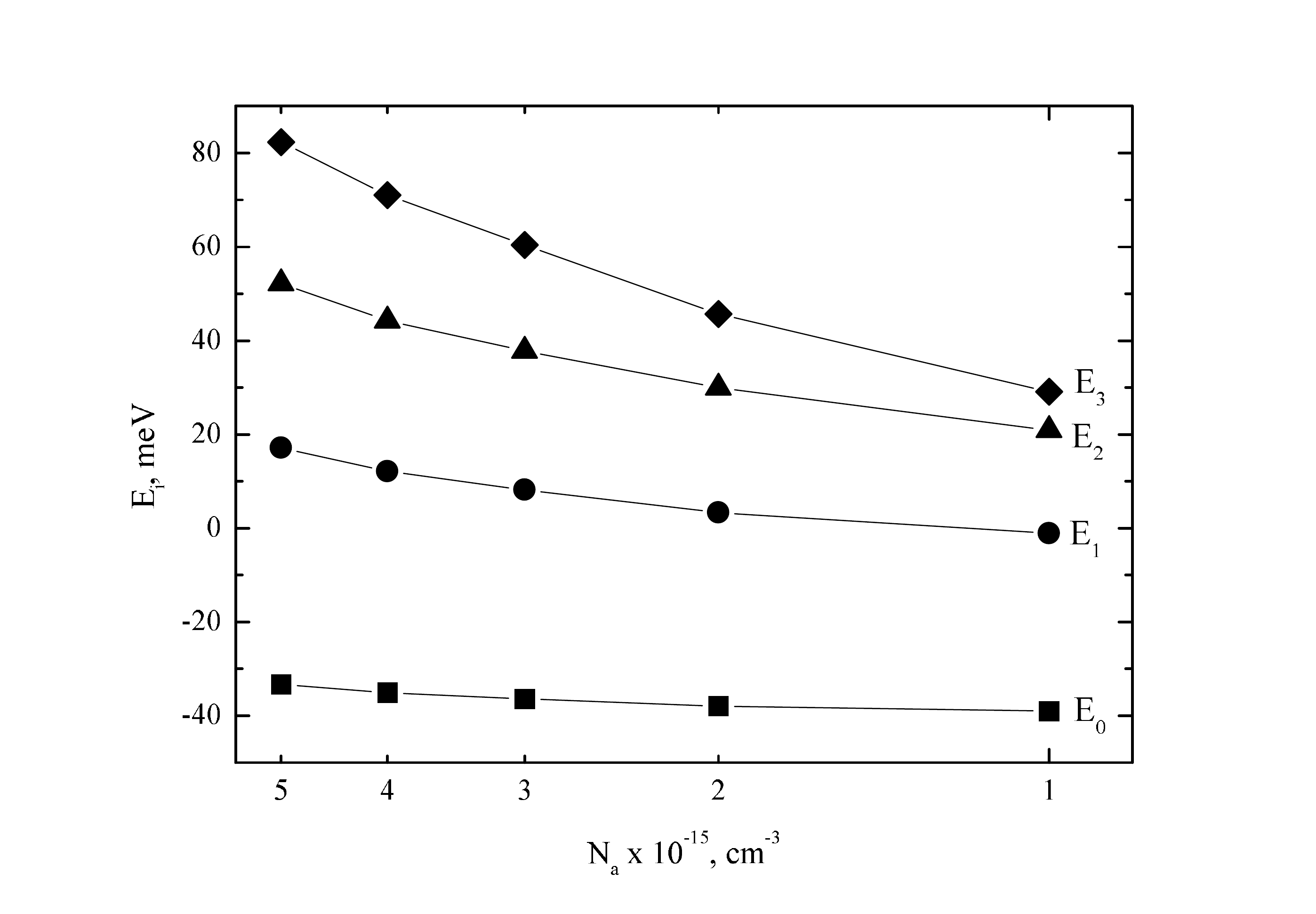}
\caption{
Theoretical calculation of the 2D-subband positions at different
acceptor concentrations at $T = 4.2$~$K$ and $U = 0$ (other
calculation parameters are the same as at the caption to the
Figure~\ref{qw}).} \label{theory}
\end{figure}

The obtained results reveal two effects in our structures at liquid helium temperatures due to illumination: the ''thickening'' of 2D levels in QW of the $\delta$-layer and the increase of the 2DEG concentration.
Let us discuss these effects.
Our experiments show that the change $\Delta\sigma_{S}$ of the tunnelling conductivity at the PTPC saturation regime does not exceed 10\% (see Figure~\ref{s_l}).
Since $\sigma$ depends on barrier height $\Phi_{S}$ at the interface Al/GaAs as $\sigma\propto\exp(-2(\Phi_{S}+E_{F})/\mathcal{E}_{00})$, the observed behaviour of $\sigma(t)$ shows negligible variation of $\Phi_{S}$ due to illumination.
Therefore, the ''thickening'' of the 2D subbands of the $\delta$-layer at PTPC regime can be explained only by the broadening of the potential well profile to the bulk GaAs direction (dashed potential profile in insert to Figure~\ref{qw}).
It can be associated with the decrease of the charged acceptors concentration $N_{a}$ in the bulk GaAs caused by the photoexiting electrons captured on acceptor levels beyond the $\delta$-doped region.
A self-consistent calculation of the dependence $E_{i}(N_{a})$ qualitatively proves this assumption (Figure~\ref{theory}).
The calculations were made at constant concentration of Si in $\delta$-layer and constant bias $U=0$ at tunnel junction.
However, the behaviour of the theoretical dependances $E_{i}(N_{a})$ reflects main features of the
experimental curves $E_{i}(t)$: subbands are ''thickening'' to $E_{0}$, and value $|E_{0}|$ is increasing by several $meV$, i.e. 2DEG concentration of the $\delta$-layer is growing due to reduction of the compensative charge on the acceptor states in bulk GaAs.

\begin{figure}[ht]
\includegraphics[width=\columnwidth]{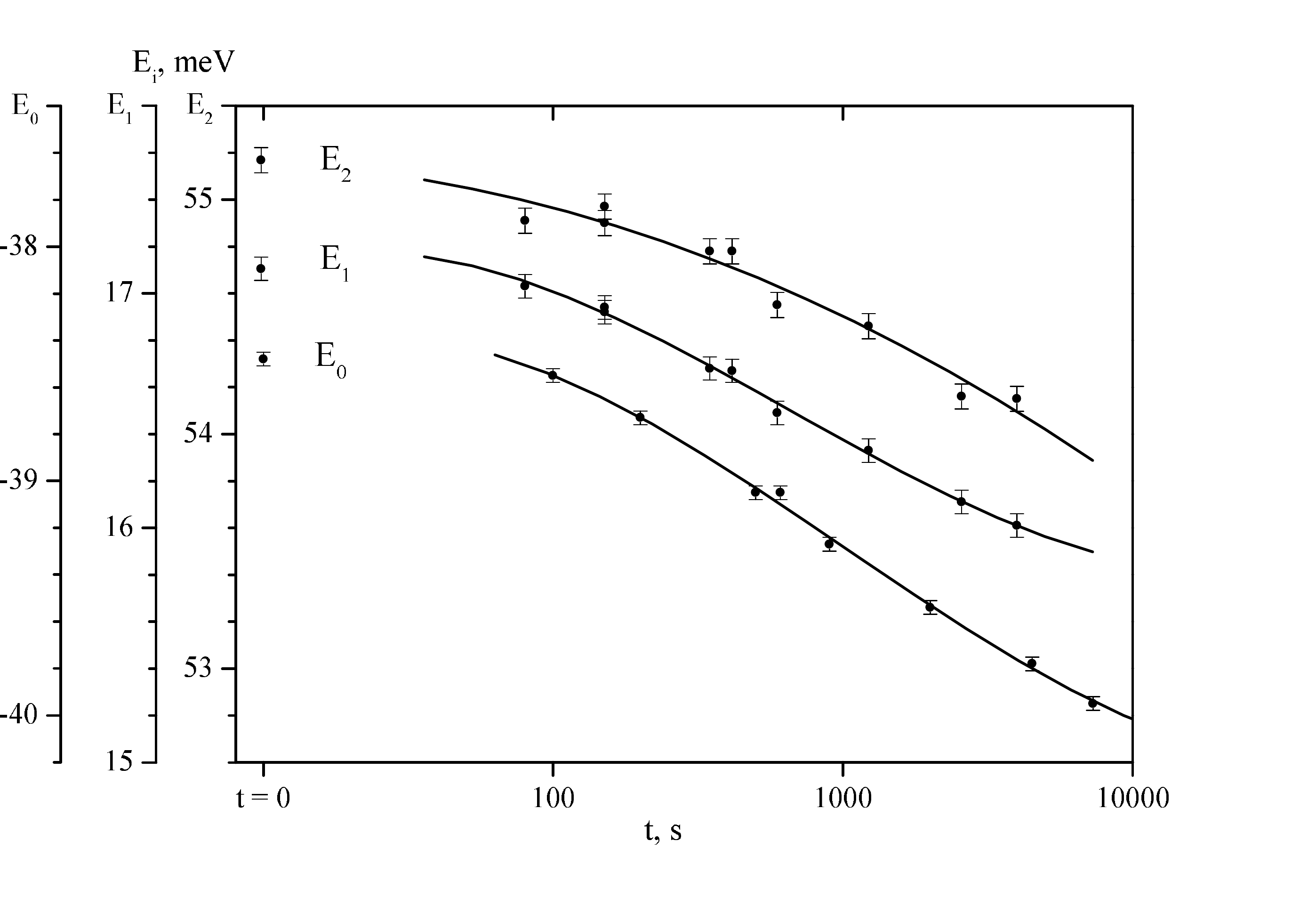}
\caption{
Persistent tunnelling photoconductivity effect at $T = 77$~$K$.
Error bars show uncertainty of measurement, solid lines are polynomial interpolations of the data.
Sample d9a11.}
\label{e_n}
\end{figure}

Note that Figure~\ref{ssnh} shows that resolution of tunnelling spectroscopy at $T=77$~$K$ is still high enough to observe 2D subbands in TS.
Accurate measurements of the TS in PTPC regime at 77~$K$ demonstrate the absence of the effect of ''thickening'' of the 2D subbands: 2D levels shift to the conduction band direction
approximately by the same value $\sim 1$~$meV$ (Figure~\ref{e_n}).
This behaviour can be associated with increase of active impurities concentration in $\delta$-layer
(because of DX-center photoionization in $\delta$-layer area,  for instance).
The model self-consistent calculations of the behaviour of $E_{i}$ as the function of the impurity concentration $N_{Si}$ in $\delta$-layer ($N_{a}=const$) demonstrate almost uniform diminution of the subband energies.
As a result the growth of $N_{Si}$ on $\simeq 2\%$ leads to level shifts by $\simeq 1$~$meV$.
And the intersubband energies practically do not change as well as experimental curves $E_{i}(t)$ at $T=77$~$K$ (Figure~\ref{e_n}).
Calculations demonstrate that this behaviour can be explained by variation of donor concentrations in $\delta$-layer.
Thus, according to tunnel spectroscopy data at 77~$K$ at PTPC regime the observed growth of 2DEG concentration is due to increase of active impurities number in $\delta$-layer only.
At liquid helium temperatures, on the contrary, the main effect of the PTPC is the change of occupied acceptors states beyond $\delta$-layer.

\begin{figure}[ht]
\includegraphics[width=\columnwidth]{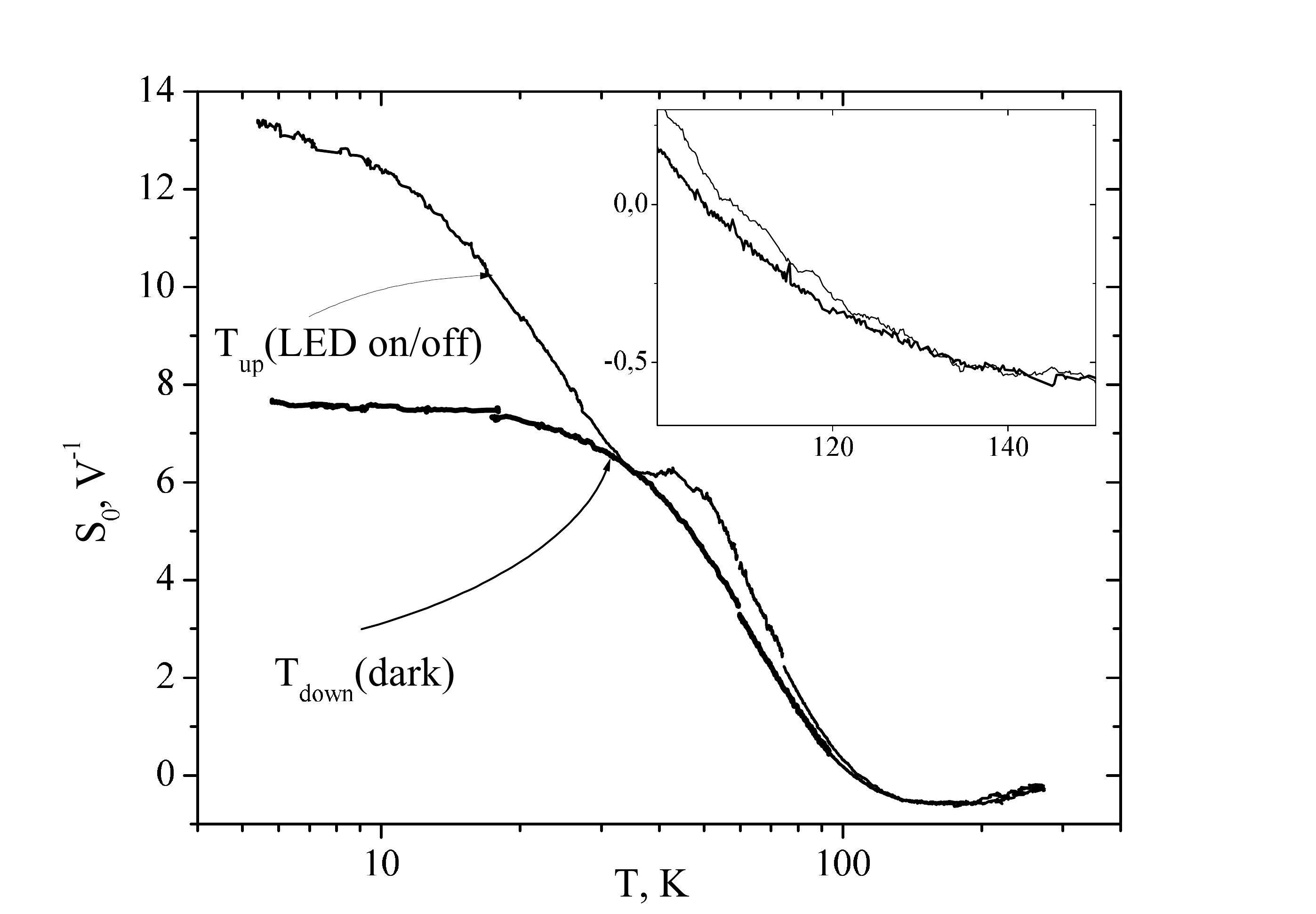}
\caption{
$S_{0}=d\ln\sigma/dU$ at $U=0$ as a function of temperature:
cooling in darkness (thick curve), reverse warming after illumination right up to saturation (thin curve).
The insert shows the behaviour of $S_{0}$ around 120~$K$.
Sample d5b42.}
\label{s_t}
\end{figure}

The data on the temperature dependence of $S_{0}=d\ln\sigma/dU$ at $U=0$ shows (Figure~\ref{s_t}) the presence of two annealing temperatures of the PTPC.
At $T\simeq 50$~$K$ the main changes in $S_{0}$ vanish. 
This apparently corresponds to thermally activation of charged centers in bulk GaAs.
The further heating of the sample suppresses change of the active donors in $\delta$-layer,  caused by illumination (at $T\ge120$~$K$ the curve $S_{0}(T)$ coincides with the ''dark'' curve).
As $T\simeq 120$~$K$ corresponds to the DX-center's activation temperature in GaAs \cite{8}, we suppose that the later effect relates to the DX-centers in the $\delta$-layer region.
The annealing temperature $T=50$~$K$ of the PPC effect in a ''bulk'' $\delta$-layer was reported in \cite{6a}, where it was related to deep centers in GaAs, which are DX-like.

\section{Conclusions}

Thus, our results show that the tunnelling spectroscopy of the near-surface delta-layer permits to obtain qualitative new information about persistent photoconductivity effect in $\delta$-layer structures on the base of GaAs.
The self-consistent and strong asymmetric potential profile (such as delta-layer QW) depends on both number of charged impurity in bulk GaAs and active donor concentration in $\delta$-layer.
Each of these factors have different effects on 2D subband's spectra in $\delta$-layer.
The decrease of acceptors concentration beyond $\delta$-layer led to ''thickening'' of the 2D subbands, whereas the increase of the active donor number in $\delta$-layer gives rise to almost uniform shift of the
dimensional quantization levels.
Note that in both above-mentioned cases the electron concentration in $\delta$-layer is augmented.
We can separate these cases through character of 2DEG spectra transformation.

According to TS data obtained, persistent effects depend upon temperature, at which the samples are illuminated.
In 50--120~$K$ temperature range  the reorganization of the TS at PTPC regime evidence of 2D electron concentration increase in QW due to donor (DX-center) photoionization in $\delta$-layer.
At temperatures less than 50~$K$ 2DEG concentration increases in addition and, besides, the spreading of the QW's potential profile of the $\delta$-layer is observed.
Additional persistent effect is associated with photoionization of the acceptor states in bulk GaAs in this case.

Besides, the results obtained show that PTPC permits to provide the resonance conditions for detecting polaron effects in the quantum well of the near-surface $\delta$-layer.

\section{Acknowledgements}
We are grateful to Yu.V. Fedorov and A.V. Hook for preparation of the samples,
V.A. Volkov for helpful discussions and
V.G. Mokerov for attention.
This work was supported by RFBR and Russian Federal Programs PAS, FTNS.

\nocite{*}
\bibliography{ptpc}

\providecommand{\noopsort}[1]{}\providecommand{\singleletter}[1]{#1}%
\begin{thebibliography}{11}%
\makeatletter
\providecommand \@ifxundefined [1]{%
 \@ifx{#1\undefined}
}%
\providecommand \@ifnum [1]{%
 \ifnum #1\expandafter \@firstoftwo
 \else \expandafter \@secondoftwo
 \fi
}%
\providecommand \@ifx [1]{%
 \ifx #1\expandafter \@firstoftwo
 \else \expandafter \@secondoftwo
 \fi
}%
\providecommand \natexlab [1]{#1}%
\providecommand \enquote  [1]{``#1''}%
\providecommand \bibnamefont  [1]{#1}%
\providecommand \bibfnamefont [1]{#1}%
\providecommand \citenamefont [1]{#1}%
\providecommand \href@noop [0]{\@secondoftwo}%
\providecommand \href [0]{\begingroup \@sanitize@url \@href}%
\providecommand \@href[1]{\@@startlink{#1}\@@href}%
\providecommand \@@href[1]{\endgroup#1\@@endlink}%
\providecommand \@sanitize@url [0]{\catcode `\\12\catcode `\$12\catcode
  `\&12\catcode `\#12\catcode `\^12\catcode `\_12\catcode `\%12\relax}%
\providecommand \@@startlink[1]{}%
\providecommand \@@endlink[0]{}%
\providecommand \url  [0]{\begingroup\@sanitize@url \@url }%
\providecommand \@url [1]{\endgroup\@href {#1}{\urlprefix }}%
\providecommand \urlprefix  [0]{URL }%
\providecommand \Eprint [0]{\href }%
\providecommand \doibase [0]{http://dx.doi.org/}%
\providecommand \selectlanguage [0]{\@gobble}%
\providecommand \bibinfo  [0]{\@secondoftwo}%
\providecommand \bibfield  [0]{\@secondoftwo}%
\providecommand \translation [1]{[#1]}%
\providecommand \BibitemOpen [0]{}%
\providecommand \bibitemStop [0]{}%
\providecommand \bibitemNoStop [0]{.\EOS\space}%
\providecommand \EOS [0]{\spacefactor3000\relax}%
\providecommand \BibitemShut  [1]{\csname bibitem#1\endcsname}%
\let\auto@bib@innerbib\@empty
\bibitem [{\citenamefont {Kotel'nikov}\ \emph {et~al.}(2000)\citenamefont
  {Kotel'nikov}, \citenamefont {Kokin}, \citenamefont {Fedorov}, \citenamefont
  {Hook},\ and\ \citenamefont {Talbaev}}]{1}%
  \BibitemOpen
  \bibfield  {author} {\bibinfo {author} {\bibfnamefont {I.~N.}\ \bibnamefont
  {Kotel'nikov}}, \bibinfo {author} {\bibfnamefont {V.~A.}\ \bibnamefont
  {Kokin}}, \bibinfo {author} {\bibfnamefont {Y.~V.}\ \bibnamefont {Fedorov}},
  \bibinfo {author} {\bibfnamefont {A.~V.}\ \bibnamefont {Hook}}, \ and\
  \bibinfo {author} {\bibfnamefont {D.~T.}\ \bibnamefont {Talbaev}},\
  }\bibfield  {title} {\enquote {\bibinfo {title} {Intersubband resonance
  polarons in al/$\delta$-gaas tunneling junctions},}\ }\href {\doibase
  10.1134/1.568361} {\bibfield  {journal} {\bibinfo  {journal} {Journal of
  Experimental and Theoretical Physics Letters}\ }\textbf {\bibinfo {volume}
  {71}},\ \bibinfo {pages} {387--390} (\bibinfo {year} {2000})}\BibitemShut
  {NoStop}%
\bibitem [{\citenamefont {A.Zrenner}(1987)}]{2a}%
  \BibitemOpen
  \bibfield  {author} {\bibinfo {author} {\bibfnamefont {A.}~\bibnamefont
  {A.Zrenner}},\ }\href@noop {} {\bibinfo {type} {{Ph.D.} thesis}},\ \bibinfo
  {school} {Technische Universität München} (\bibinfo {year}
  {1987})\BibitemShut {NoStop}%
\bibitem [{\citenamefont {Koenraad}\ \emph {et~al.}(1990)\citenamefont
  {Koenraad}, \citenamefont {Voncken}, \citenamefont {Singleton}, \citenamefont
  {Blom}, \citenamefont {Langerak}, \citenamefont {Leys}, \citenamefont
  {Perenboom}, \citenamefont {Spermon}, \citenamefont {{van Vleuten}},\ and\
  \citenamefont {Wolter}}]{2}%
  \BibitemOpen
  \bibfield  {author} {\bibinfo {author} {\bibfnamefont {P.}~\bibnamefont
  {Koenraad}}, \bibinfo {author} {\bibfnamefont {A.}~\bibnamefont {Voncken}},
  \bibinfo {author} {\bibfnamefont {J.}~\bibnamefont {Singleton}}, \bibinfo
  {author} {\bibfnamefont {F.}~\bibnamefont {Blom}}, \bibinfo {author}
  {\bibfnamefont {C.}~\bibnamefont {Langerak}}, \bibinfo {author}
  {\bibfnamefont {M.}~\bibnamefont {Leys}}, \bibinfo {author} {\bibfnamefont
  {J.}~\bibnamefont {Perenboom}}, \bibinfo {author} {\bibfnamefont
  {S.}~\bibnamefont {Spermon}}, \bibinfo {author} {\bibfnamefont
  {W.}~\bibnamefont {{van Vleuten}}}, \ and\ \bibinfo {author} {\bibfnamefont
  {J.}~\bibnamefont {Wolter}},\ }\bibfield  {title} {\enquote {\bibinfo {title}
  {Characterization of silicon $\delta$-doped gaas grown by mbe at various
  temperatures},}\ }\href {\doibase
  https://doi.org/10.1016/0039-6028(90)90371-E} {\bibfield  {journal} {\bibinfo
   {journal} {Surface Science}\ }\textbf {\bibinfo {volume} {228}},\ \bibinfo
  {pages} {538 -- 541} (\bibinfo {year} {1990})}\BibitemShut {NoStop}%
\bibitem [{\citenamefont {Arscott}, \citenamefont {Missous},\ and\
  \citenamefont {Dobaczewski}(1992)}]{3}%
  \BibitemOpen
  \bibfield  {author} {\bibinfo {author} {\bibfnamefont {S.}~\bibnamefont
  {Arscott}}, \bibinfo {author} {\bibfnamefont {M.}~\bibnamefont {Missous}}, \
  and\ \bibinfo {author} {\bibfnamefont {L.}~\bibnamefont {Dobaczewski}},\
  }\bibfield  {title} {\enquote {\bibinfo {title} {Observation of persistent
  photoconductivity in delta-doped {GaAs}},}\ }\href {\doibase
  10.1088/0268-1242/7/4/033} {\bibfield  {journal} {\bibinfo  {journal}
  {Semiconductor Science and Technology}\ }\textbf {\bibinfo {volume} {7}},\
  \bibinfo {pages} {620--623} (\bibinfo {year} {1992})}\BibitemShut {NoStop}%
\bibitem [{\citenamefont {de~Oliveira}\ \emph {et~al.}(1995)\citenamefont
  {de~Oliveira}, \citenamefont {Ribeiro}, \citenamefont {Soares}, \citenamefont
  {Corrêa~F.}, \citenamefont {da~Silva},\ and\ \citenamefont {Chacham}}]{4}%
  \BibitemOpen
  \bibfield  {author} {\bibinfo {author} {\bibfnamefont {A.~G.}\ \bibnamefont
  {de~Oliveira}}, \bibinfo {author} {\bibfnamefont {G.~M.}\ \bibnamefont
  {Ribeiro}}, \bibinfo {author} {\bibfnamefont {D.~A.~W.}\ \bibnamefont
  {Soares}}, \bibinfo {author} {\bibfnamefont {J.~A.}\ \bibnamefont
  {Corrêa~F.}}, \bibinfo {author} {\bibfnamefont {M.~I.~N.}\ \bibnamefont
  {da~Silva}}, \ and\ \bibinfo {author} {\bibfnamefont {H.}~\bibnamefont
  {Chacham}},\ }\bibfield  {title} {\enquote {\bibinfo {title}
  {Photoconductivity and n‐ to p‐type transition in silicon planar‐doped
  gaas structures with a doped cap layer},}\ }\href {\doibase 10.1063/1.360127}
  {\bibfield  {journal} {\bibinfo  {journal} {Journal of Applied Physics}\
  }\textbf {\bibinfo {volume} {78}},\ \bibinfo {pages} {2659--2665} (\bibinfo
  {year} {1995})},\ \Eprint
  {http://arxiv.org/abs/https://doi.org/10.1063/1.360127}
  {https://doi.org/10.1063/1.360127} \BibitemShut {NoStop}%
\bibitem [{\citenamefont {Li}\ and\ \citenamefont {Jagadish}(1997)}]{4a}%
  \BibitemOpen
  \bibfield  {author} {\bibinfo {author} {\bibfnamefont {G.}~\bibnamefont
  {Li}}\ and\ \bibinfo {author} {\bibfnamefont {C.}~\bibnamefont {Jagadish}},\
  }\bibfield  {title} {\enquote {\bibinfo {title} {Effect of illumination on
  the subband electronic structure of si $\delta$-doped gaas},}\ }\href
  {\doibase 10.1063/1.119317} {\bibfield  {journal} {\bibinfo  {journal}
  {Applied Physics Letters}\ }\textbf {\bibinfo {volume} {70}},\ \bibinfo
  {pages} {90--92} (\bibinfo {year} {1997})},\ \Eprint
  {http://arxiv.org/abs/https://doi.org/10.1063/1.119317}
  {https://doi.org/10.1063/1.119317} \BibitemShut {NoStop}%
\bibitem [{\citenamefont {Valyaev}\ \emph {et~al.}(1998)\citenamefont
  {Valyaev}, \citenamefont {Gurtovoi}, \citenamefont {Ivanov}, \citenamefont
  {Morozov}, \citenamefont {Sirotkin}, \citenamefont {Dubrovskii},
  \citenamefont {Shapoval}, \citenamefont {Khanin}, \citenamefont {Vdovin},\
  and\ \citenamefont {Pustovit}}]{5}%
  \BibitemOpen
  \bibfield  {author} {\bibinfo {author} {\bibfnamefont {V.~V.}\ \bibnamefont
  {Valyaev}}, \bibinfo {author} {\bibfnamefont {V.~L.}\ \bibnamefont
  {Gurtovoi}}, \bibinfo {author} {\bibfnamefont {D.~Y.}\ \bibnamefont
  {Ivanov}}, \bibinfo {author} {\bibfnamefont {S.~V.}\ \bibnamefont {Morozov}},
  \bibinfo {author} {\bibfnamefont {V.~V.}\ \bibnamefont {Sirotkin}}, \bibinfo
  {author} {\bibfnamefont {Y.~V.}\ \bibnamefont {Dubrovskii}}, \bibinfo
  {author} {\bibfnamefont {S.~Y.}\ \bibnamefont {Shapoval}}, \bibinfo {author}
  {\bibfnamefont {Y.~N.}\ \bibnamefont {Khanin}}, \bibinfo {author}
  {\bibfnamefont {E.~E.}\ \bibnamefont {Vdovin}}, \ and\ \bibinfo {author}
  {\bibfnamefont {A.~N.}\ \bibnamefont {Pustovit}},\ }\bibfield  {title}
  {\enquote {\bibinfo {title} {Conductivity and persistent photoconductivity in
  gaas epitaxial films containing single and double delta-doped layers},}\
  }\href {\doibase 10.1134/1.558440} {\bibfield  {journal} {\bibinfo  {journal}
  {Journal of Experimental and Theoretical Physics}\ }\textbf {\bibinfo
  {volume} {86}},\ \bibinfo {pages} {383--387} (\bibinfo {year}
  {1998})}\BibitemShut {NoStop}%
\bibitem [{\citenamefont {Chen}\ \emph {et~al.}(1998)\citenamefont {Chen},
  \citenamefont {Thio}, \citenamefont {Wang}, \citenamefont {Alt},\ and\
  \citenamefont {Sharma}}]{6a}%
  \BibitemOpen
  \bibfield  {author} {\bibinfo {author} {\bibfnamefont {C.~Y.}\ \bibnamefont
  {Chen}}, \bibinfo {author} {\bibfnamefont {T.}~\bibnamefont {Thio}}, \bibinfo
  {author} {\bibfnamefont {K.~L.}\ \bibnamefont {Wang}}, \bibinfo {author}
  {\bibfnamefont {K.~W.}\ \bibnamefont {Alt}}, \ and\ \bibinfo {author}
  {\bibfnamefont {P.~C.}\ \bibnamefont {Sharma}},\ }\bibfield  {title}
  {\enquote {\bibinfo {title} {Persistent photoconductivity in si delta-doped
  gaas at low doping concentration},}\ }\href {\doibase 10.1063/1.122729}
  {\bibfield  {journal} {\bibinfo  {journal} {Applied Physics Letters}\
  }\textbf {\bibinfo {volume} {73}},\ \bibinfo {pages} {3235--3237} (\bibinfo
  {year} {1998})},\ \Eprint
  {http://arxiv.org/abs/https://doi.org/10.1063/1.122729}
  {https://doi.org/10.1063/1.122729} \BibitemShut {NoStop}%
\bibitem [{\citenamefont {Piotrzkowski}\ \emph {et~al.}(1999)\citenamefont
  {Piotrzkowski}, \citenamefont {Litwin-Staszewska}, \citenamefont {Bosc},
  \citenamefont {Sicart},\ and\ \citenamefont {Robert}}]{6}%
  \BibitemOpen
  \bibfield  {author} {\bibinfo {author} {\bibfnamefont {R.}~\bibnamefont
  {Piotrzkowski}}, \bibinfo {author} {\bibfnamefont {E.}~\bibnamefont
  {Litwin-Staszewska}}, \bibinfo {author} {\bibfnamefont {F.}~\bibnamefont
  {Bosc}}, \bibinfo {author} {\bibfnamefont {J.}~\bibnamefont {Sicart}}, \ and\
  \bibinfo {author} {\bibfnamefont {J.}~\bibnamefont {Robert}},\ }\bibfield
  {title} {\enquote {\bibinfo {title} {New type of persistent photoconductivity
  related to dx-center: the study of interband ppc in si-doped algaas},}\
  }\href {\doibase https://doi.org/10.1016/S0921-4526(99)00651-1} {\bibfield
  {journal} {\bibinfo  {journal} {Physica B: Condensed Matter}\ }\textbf
  {\bibinfo {volume} {273-274}},\ \bibinfo {pages} {792 -- 795} (\bibinfo
  {year} {1999})}\BibitemShut {NoStop}%
\bibitem [{7(2001)}]{7}%
  \BibitemOpen
  \href@noop {} {\emph {\bibinfo {title} {Proc. of All-Russian Workshop
  ”Nanophotonics”}}}\ (\bibinfo {organization} {IPM RAS},\ \bibinfo
  {address} {N.Novgorod},\ \bibinfo {year} {2001})\ \bibinfo {note} {in
  Russian}\BibitemShut {NoStop}%
\bibitem [{\citenamefont {Mooney}(1990)}]{8}%
  \BibitemOpen
  \bibfield  {author} {\bibinfo {author} {\bibfnamefont {P.~M.}\ \bibnamefont
  {Mooney}},\ }\bibfield  {title} {\enquote {\bibinfo {title} {Deep donor
  levels (dx centers) in iii‐v semiconductors},}\ }\href {\doibase
  10.1063/1.345628} {\bibfield  {journal} {\bibinfo  {journal} {Journal of
  Applied Physics}\ }\textbf {\bibinfo {volume} {67}},\ \bibinfo {pages}
  {R1--R26} (\bibinfo {year} {1990})},\ \Eprint
  {http://arxiv.org/abs/https://doi.org/10.1063/1.345628}
  {https://doi.org/10.1063/1.345628} \BibitemShut {NoStop}%
\end{thebibliography}%


\providecommand{\noopsort}[1]{}\providecommand{\singleletter}[1]{#1}%
%

\end{document}